\documentstyle[12pt,epsfig,rotating]{article}
\newcommand{\kjet}{k_{T{\rm jet}}}
\newcommand{\xjet}{x_{\rm jet}}
\newcommand{\Ejet}{E_{\rm jet}}
\newcommand{\thjet}{\theta_{\rm jet}}
\newcommand{\pjet}{p_{T{\rm jet}}}

\newcommand{\xB}{\mbox{$x~$}}  

\newcommand{\Qsq}{\mbox{$Q^2~$}}

\newcommand{\GeVx}{\rm GeV}

\newcommand{\GeVsq}{\mbox{${\rm ~GeV}^2~$}}

\newcommand {\gapprox}
   {\raisebox{-0.7ex}{$\stackrel {\textstyle>}{\sim}$}}

\def\gsim{\,\lower.25ex\hbox{$\scriptstyle\sim$}\kern-1.30ex%
\raise 0.55ex\hbox{$\scriptstyle >$}\,}
\def\lsim{\,\lower.25ex\hbox{$\scriptstyle\sim$}\kern-1.30ex%
\raise 0.55ex\hbox{$\scriptstyle <$}\,}

\newcommand{\piz}{$\pi^{0}$}

\newcommand{\etpi}{\mbox{$E_{T,\pi}$}}

\newcommand{\epi}{\mbox{$E_{\pi}$}}
\newcommand{\ecp}{\mbox{$E_{\mathrm{cp}}$}}
\newcommand{\xpi}{\mbox{$x_{\pi}$}}
\newcommand{\xcp}{\mbox{$x_{\mathrm{cp}}$}}

\newcommand{\thpi}{\mbox{$\theta_{\pi}$}}
\newcommand{\thcp}{\mbox{$\theta_{\mathrm{cp}}$}}

\newlength{\dinwidth}
\newlength{\dinmargin}
\begin{document}
\setlength{\unitlength}{1mm}
\begin{titlepage}
\begin{flushleft}
%
%
{\tt DESY 98-143    \hfill    ISSN 0418-9833} \\
{\tt September 1998}                  \\
\end{flushleft}
\vspace*{4.cm}
\begin{center}
\begin{Large}
 
{\boldmath \bf Forward Jet and Particle Production at HERA }\\
 
\vspace{1.cm}
{H1 Collaboration}    \\
\end{Large} \end{center}
\vspace*{2.5cm}
\begin{abstract} 
\noindent Single particles and jets in deeply inelastic scattering
 at low $x$ are measured with the H1 detector
in the region away from   
   the current jet and towards 
the proton remnant, known as the forward region. Hadronic final state
measurements in this region are  
expected to be particularly sensitive to  QCD 
evolution effects. 
Jet cross-sections 
are presented  as a function of Bjorken-$x$ 
for forward jets produced with a polar angle to the proton
direction, $\thjet$, in the range $7^{\circ}< \thjet<20^{\circ}$.  Azimuthal
correlations are studied between the forward jet and the scattered lepton.
Charged and neutral single particle production in the forward region 
are measured  as a function of Bjorken-$x$,
in the range 
$5^{\circ}< \theta      <25^{\circ}$, 
for particle transverse momenta  larger than 1 GeV.
QCD based Monte Carlo predictions 
and analytical calculations based  
on BFKL, CCFM and DGLAP evolution 
are compared to the data. Predictions based on the 
DGLAP approach fail to describe the data, except for those which allow
for a resolved photon contribution.

\end{abstract}
%
%
%
\vspace{2.cm}
\hspace{5.0cm} {\it Submitted to Nuclear Physics B }  
%
\end{titlepage}
\begin{Large} \begin{center} H1 Collaboration \end{center} \end{Large}
\begin{flushleft}
 C.~Adloff$^{34}$,                
 M.~Anderson$^{22}$,              
 V.~Andreev$^{25}$,               
 B.~Andrieu$^{28}$,               
 V.~Arkadov$^{35}$,               
 C.~Arndt$^{11}$,                 
 I.~Ayyaz$^{29}$,                 
 A.~Babaev$^{24}$,                
 J.~B\"ahr$^{35}$,                
 J.~B\'an$^{17}$
% ...

\end{flushleft}
\begin{flushleft} {\it
% ...
$ ^{11}$ DESY, Hamburg, Germany$^a$ \\
% ...
$ ^{17}$ Institute of Experimental Physics, Slovak Academy of
          Sciences, Ko\v{s}ice, Slovak Republic$^{f,j}$ \\
% ...
$ ^{22}$ Department of Physics and Astronomy, 
          University of Manchester, Manchester, UK$^b$ \\
% ...
$ ^{24}$ Institute for Theoretical and Experimental Physics,
          Moscow, Russia \\
 $ ^{25}$ Lebedev Physical Institute, Moscow, Russia$^{f,k}$ \\
% ...
$ ^{28}$ LPNHE, \'{E}cole Polytechnique, IN2P3-CNRS, Palaiseau, France \\
 $ ^{29}$ LPNHE, Universit\'{e}s Paris VI and VII, IN2P3-CNRS,
          Paris, France \\
% ...
$ ^{34}$ Fachbereich Physik, Bergische Universit\"at Gesamthochschule
          Wuppertal, Wuppertal, Germany$^a$ \\
 $ ^{35}$ DESY, Institut f\"ur Hochenergiephysik, Zeuthen, Germany$^a$ \\
% ...
$ ^a$ Supported by the Bundesministerium f\"ur Bildung, Wissenschaft,
        Forschung und Technologie, FRG,
        under contract numbers 7AC17P, 7AC47P, 7DO55P, 7HH17I, 7HH27P,
        7HD17P, 7HD27P, 7KI17I, 6MP17I and 7WT87P \\
 $ ^b$ Supported by the UK Particle Physics and Astronomy Research
       Council, and formerly by the UK Science and Engineering Research
       Council \\
% ...
$ ^f$ Supported by the Deutsche Forschungsgemeinschaft \\
% ...
$ ^j$ Supported by VEGA SR grant no. 2/5167/98 \\
 $ ^k$ Supported by Russian Foundation for Basic Research 
       grant no. 96-02-00019 

   } \end{flushleft}
 
%
\newpage
\section{Introduction}
\noindent
 
\begin{figure}[htb] \centering \unitlength 1mm
\epsfig{file=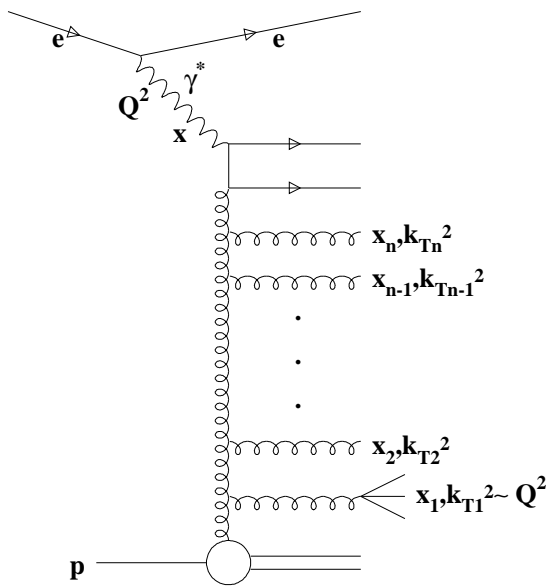,height=11cm}
\caption{Parton evolution in the ladder approximation. The
kinematics of forward jets in DIS events are indicated.}
\label{HOTSPOT}
\end{figure}

The 
electron-proton collider HERA has made possible
measurements of deeply inelastic scattering (DIS) in new kinematic regions:
 the regions of large
four-momentum 
transfer \Qsq (up to $\Qsq \approx 10^4 \GeVsq$) and  small Bjorken-\xB
(\xB $ \approx 10^{-4}$). A diagram illustrating the QCD evolution of 
a low $x$ DIS event in the infinite momentum frame
is shown in Fig.~\ref{HOTSPOT};    a parton in the proton
 is able to undergo a QCD cascade resulting in several parton emissions 
before the final parton interacts with the virtual photon. These partons 
convert into hadrons which can then be detected and measured. 
The measurement of the hadronic final state between the 
proton remnant and  the struck quark  has been advocated over
the past years  as 
a technique
for studying the dynamics of the QCD cascade in small $x$ 
DIS~\cite{mueller,dhotref,hotref}.

In this analysis data from the H1 experiment are used to study 
single particles and jets in the forward region, defined as
 polar angle, $\theta$, 
less than 25$^ \circ$ in the laboratory reference frame    
measured with respect to the proton beam direction. 
A complication in using the hadronic final state to 
investigate QCD evolution arises due to the soft hadronisation 
processes. It has been suggested that this can be reduced by 
 studying   
 forward jet 
production~\cite{mueller,dhotref,hotref}.  
Such an analysis, 
based on a data sample containing approximately 
one tenth of the  statistics used in this study, was reported 
in~\cite{forwjets}. A similar analysis has been reported 
recently in~\cite{fjzeus}.
Single high-$p_T$ particle production as a test of the underlying QCD dynamics 
has also been proposed~\cite{kuhlen}. A first analysis of  
transverse momentum ($p_T$) and pseudorapidity ($\eta$) spectra was made
using charged particles in the pseudorapidity region $0.5<\eta<2.5$
in the $\gamma^* p$ centre of mass frame~\cite{charged}, 
accessing  particles with angles  down to  $8^{\circ}$ in the 
laboratory frame. 
The advantages of studying single particles, rather than jets
 lie in their independence of any choice of jet finding algorithm 
and also the potential to reach smaller angles 
than would be possible with jets with broad spatial extent.
For single particles however, fragmentation effects are expected to be 
more significant. Thus both measurements are complementary.

Several prescriptions for the QCD dynamics in the region of phase space 
towards the proton remnant have been proposed and are compared to the H1 
data in this study. These include QCD parton evolution schemes such as 
the classical 
DGLAP (Dokshitzer-Gribov-Lipatov-Altarelli-Parisi)~\cite{dglap}
evolution, the low $x$ specific BFKL
(Balitsky-Fadin-Kuraev-Lipatov)~\cite{bfkl} evolution equation and
the CCFM
(Ciafaloni-Catani-Fiorani-Marchesini)~\cite{ccfm} evolution 
equation, which  
forms a bridge between 
the BFKL and DGLAP approaches. 
When the squared transverse momentum of a parton emitted in the ladder
(Fig.\ref{HOTSPOT})
is larger than  $Q^2$ the resolved photon picture becomes applicable,
in which case
 the photon can be considered 
to interact via its own resolved partonic
component. 

\section{QCD calculations and phenomenology}

DGLAP 
evolution, in which the relevant evolution parameter is $\ln Q^2$,  
 has been successfully tested in large $Q^2$ processes 
and provides an excellent  description of the scale dependence of the structure function 
of the proton. 
 The BFKL
 evolution equation, for which
the relevant evolution parameter is $\ln1/x$, 
 describes scattering processes in QCD in the 
limit of large energies and fixed, but sufficiently large, momentum transfers.
At lowest order the BFKL and DGLAP  evolution equations effectively resum the 
leading logarithmic contributions  $\alpha_s\ln1/x$ and 
$\alpha_s\ln Q^2$  respectively which, in an axial gauge,
amounts to a resummation of ladder diagrams of the type shown in
Fig.~\ref{HOTSPOT}. Since 
it is expected that at small enough $x$ the 
$\ln1/x$ terms dominate the evolution, the BFKL equation
should be applicable in this region.
In the leading log DGLAP scheme 
the parton cascade follows a strong ordering in transverse
momenta 
$k_{Tn}^2 \gg k_{Tn-1}^2 \gg... \gg k_{T1}^2$, 
while there is only a soft 
(kinematic) ordering for the fractional momenta $x_n<x_{n-1}<...<x_1$.
In the BFKL scheme, the cascade follows a strong ordering in fractional
momentum  
$x_n \ll x_{n-1} \ll... \ll x_1$, 
while there is no ordering in transverse
momentum~\cite{muellercarg}.
 The transverse momenta perform a kind of random walk 
in $k_{T}$ space ; 
 $k_{Ti}$ is close to $k_{Ti-1}$, but it
can be both larger or smaller~\cite{bartels}. For the  CCFM
 evolution 
equation 
the relevant quantity is the angular ordering in the cascade.
 
In a model that includes the resolved photon component in DIS
 the photon can be considered 
to interact via its own resolved partonic
component. 
This internal photon structure leads to 
 parton-parton scattering that can give rise 
to high $E_T$ jets, and emulate a random walk 
behaviour in $k_T$~\cite{jung}. This is usually theoretically treated by
ascribing a structure function to the photon and allowing DGLAP evolution
to take place from partons in both the photon and the proton.
The ladder shown in Fig.\ref{HOTSPOT} will then consist of two smaller
ladders, one between the hard scattering and the photon, and one between
the hard scattering and the proton. 
Consequently there is no longer a strict ordering
in $k_T$ from the photon to the proton vertex, as 
obtained using the DGLAP picture without allowing virtual photon structure.

Predictions for final state observables are available 
from Monte Carlo models, based upon QCD phenomenology.
LEPTO 6.5~\cite{lepto} incorporates first order QCD matrix elements 
with matched leading-log DGLAP parton showers to simulate higher-order 
radiation (MEPS). For this model, the factorisation and renormalisation 
scales are set to $Q^2$.  
The MEPS scheme 
fails to account for the di-jet rates when $E_{T{\rm jet}}^2  \gapprox
Q^2$~\cite{tania,h1dijets} in the kinematic range studied in this paper,
although this is not unexpected since for 
 $E_{T{\rm jet}}^2 \gapprox Q^2$ the parton from the proton is able to 
probe the structure of the photon. 
Using the  concept of a resolved photon, as suggested above,  
the measured di-jet rates can be described~\cite{jung,h1dijets}. 
 The Monte Carlo RAPGAP 2.06~\cite{rapgap} can include such a 
 resolved photon contribution in addition to the  standard 
``direct" process.
For the study presented here  the  
 SAS-2D parametrisation~\cite{SAS1D}  of the virtual photon parton
densities was used, and $Q^2+p_T^2$, with 
$p_T$ the transverse momentum of the partons in the 
hard scattering,  was used for the factorisation
and renormalisation scales.

 ARIADNE 4.08~\cite{ariadne}
provides an implementation of the Colour Dipole
Model (CDM) of a chain of independently radiating dipoles formed by
emitted gluons~\cite{dipole}. 
Unlike that in LEPTO, the parton cascade generated in the CDM is not 
 $k_T$ ordered~\cite{bfklcdm}. 
Since all radiation is
assumed to  come from the dipole
formed by the struck quark and the remnant, photon-gluon fusion events
have to be added and are taken from the QCD matrix elements.
 The linked dipole chain (LDC) model~\cite{ldc} is a reformulation of the 
CCFM equation and redefines the separation of initial and final state 
QCD emissions using the CDM. This overcomes
technical problems in calculating non-Sudakov form factors for
 parton splitting functions. Calculations of the hadronic final 
state based on this approach are available within the LDCMC  
1.0~\cite{Lonnblad-ldc}
Monte Carlo which matches exact first order matrix elements with  
the LDC-prescribed initial and final state parton emissions. 
Both the LDCMC and ARIADNE use $p_T^2$ as the factorisation and 
renormalisation scales. 

 For the 
hadronisation of the partonic system,  all of the above models 
 use the Lund string model as implemented in JETSET~\cite{string}. 
 All of these models are used with the  GRV-HO~\cite{GRV}
proton parton densities. 
QED corrections are determined with the Monte Carlo program 
DJANGO~\cite{django}, using the CDM for the final state.

      Alternatively,
        analytical BFKL calculations~\cite{adrbartels,kwiecinski} 
for forward jet production
are available 
 at the parton level.
  These calculations are 
based on asymptotic expressions derived from the BFKL equation 
at leading order (LO). 
 In order to compare to 
single particle spectra a different approach has been adopted.
The normalization of the calculation was fixed by comparing the 
prediction for the forward jet cross sections with the data,
as reported in~\cite{kwiecinski}. Once this normalization 
was fixed, predictions for particle spectra were made for the 
kinematic range given below, using 
 fragmentation
functions.

Also available are exact calculations 
 of jet cross-sections
 at the parton level  ,
at fixed 
order ${\alpha}_s^2$  in the strong coupling
constant,  for direct or pointlike interactions 
of the photon with the proton,  as  implemented in the 
DISENT~\cite{disent} package.
 A comparison with DISENT for di-jet cross-sections 
 has been reported in~\cite{h1dijets}.

Background from photoproduction processes, those events for which 
$Q^2 \sim 0$ 
in which  the electron
  remains undetected in the beampipe and
a fake electron is detected in the hadronic final state, has been 
studied using the PHOJET~\cite{phojet} Monte Carlo program. This 
generator contains 
LO QCD matrix elements for hard subprocesses, 
a  parton showers model and a 
phenomenological description of  soft processes.

\section{The H1 detector}
A detailed description of the H1 apparatus  
can be found elsewhere~\cite{h1nim}.
The following section briefly describes the components of the detector
relevant 
to this 
analysis. 
 
The hadronic energy flow and the scattered electron are measured with a
liquid argon~(LAr) calorimeter and a  
backward electromagnetic lead-scintillator calorimeter (BEMC) respectively. 
The LAr calorimeter~\cite{larc}  extends over the polar angle range
$4^\circ < \theta <  154^\circ$ with full azimuthal coverage. 
It consists of an electromagnetic section with lead absorbers
and a hadronic section with steel absorbers.
Both sections are highly segmented in the transverse and 
the longitudinal direction, in particular in the forward region of the
detector,
having about $45\,000$ channels in total.
The total depth of both calorimeters 
varies between 4.5 and 8 interaction lengths in the region $ 4^\circ <
\theta < 128^\circ$.
Test beam measurements of the LAr~calorimeter modules show an
energy resolution 
of $\sigma_{E}/E\approx 0.50/\sqrt{E\;[\GeVx]} \oplus 0.02$  for 
charged pions~\mbox{\cite{h1pi}}. 
The hadronic energy measurement is performed 
by applying a weighting technique in
order to account 
for the non-compensating nature of the calorimeter.
The 
absolute scale of the hadronic energy is presently known to $4\%$, as
determined from studies of the 
transverse momentum ($p_T$) balance in DIS events 
with forward jets~\cite{ewelina}.
 
The BEMC (depth 22 radiation lengths or 1
interaction length) covers the backward region of the detector,
$151^\circ < \theta < 176^\circ$.
The primary task  of the BEMC is to trigger on 
   DIS processes with $Q^2$ values ranging from 5 to 100 GeV$^{2}$,
and to measure the
scattered electron in DIS events. 
For scattered electrons with an energy larger than
11 GeV, as used in this analysis, the trigger efficiency is better 
than 99\%.
The BEMC energy scale for electrons is known to an accuracy of 
$1\%$~\cite{f2pap}.
Its resolution is given by 
$\sigma_{E}/E = 0.10/\sqrt{E\;[\GeVx]} \oplus 0.39/E[\GeVx] \oplus 0.017$
\cite{bemcnim}.
 
The calorimeters are 
surrounded by a superconducting solenoid providing a uniform
magnetic field of $1.15$ T parallel to the beam axis in the tracking region.
Charged particle 
tracks are measured in the central tracker (CT)
covering the polar angular range $ 20^\circ < \theta < 160^\circ$ and
 the forward 
tracking (FT) system,
covering the polar angular range $ 5^\circ < \theta < 25^\circ$. 
The CT consists of inner and outer cylindrical jet chambers, $z$-drift 
chambers and proportional chambers.  The jet chambers,  mounted
concentrically around the beam line, cover  
a range of polar angle of  $ 20^\circ < \theta < 165^\circ$ and  
maximally provide  65 space points in the radial plane
for tracks with sufficiently large transverse momentum. 
The achieved resolutions are 
$\sigma_{p_T}/p_T \approx 0.009\cdot p_T\;[\GeVx] \oplus 0.015$
and $\sigma_\theta = 20$~mrad \cite{h1nim}.
The FT provides measurements of tracks in the range  
$5^\circ < \theta <25^\circ$. This device has  three identical 
sections, each containing a series of proportional, and 
radial and planar drift chambers,
specifically arranged to facilitate triggering and
track reconstruction at low 
polar angle. For tracks fitted to the interaction vertex the resolution 
 in $p_T$ has been demonstrated to be $\sigma_{p_T}/p_T \approx 0.02\cdot
p_T\;[\GeVx] \oplus 0.1$ and the angular resolution,  $\sigma_\theta$, 
to be better than 1 mrad~\cite{FT}.   

A backward proportional chamber (BPC), in front of the BEMC with an angular
acceptance of $155.5^\circ < \theta < 174.5^\circ$ serves to identify electron
candidates
and to precisely measure their direction. 
Using information from the BPC, the BEMC and the reconstructed event vertex the
polar angle of the scattered electron is known to about 1 mrad. 


The luminosity is measured using the 
reaction $ep\rightarrow ep\gamma$
with  two TlCl/TlBr crystal calorimeters, installed
in the HERA tunnel.
The electron tagger is located at $z=-33$ m and the photon tagger 
at $z=-103$ m from the interaction point in the direction of the outgoing
electron beam.

\section{Data selection and corrections}{\label{sec:datasel}}

Experimental data for this analysis were collected by H1 during the 1994
running period, in which HERA collided 27.5 GeV 
electrons\footnote{In 1994 the incident 
lepton at HERA was a positron although we use the generic name
`electron' for the incident and scattered lepton throughout this paper.}
 with 820 GeV
protons. Integrated luminosities of the data samples used for the jets,
 $\pi^0$ and charged particle measurements are 2.8, 2.0 and 1.2  pb$^{-1}$ 
respectively. These differences arise due to the importance of selecting 
running periods with optimal experimental conditions  for
each of the measurements presented here.

 DIS events are selected~\cite{f2pap} via
the scattered electron
which is experimentally defined as a 
 high energy cluster, i.e. a localised  
energy deposit in the BEMC, 
with a cluster radius less than 
5 cm and with an associated hit in the 
BPC. The scattered electron is required to satisfy
 $E_{e}^{'} > 12 $ GeV and $\theta_{min} = 156^{\circ}$ 
(for the particle analysis), 
or $E_{e}^{'}> 11 $ GeV and $\theta_{min} = 160^{\circ}$ 
(for the jet analysis), 
$\theta_{min} < \theta_{e} < 173^{\circ}$ 
where $E_e^{'}$
and $\theta_{e}$ are the energy and angle of the scattered electron.
All results quoted here are corrected to their respective 
range.
The kinematic variables are determined using information from the
scattered electron:
  $ Q^{2} = 4 E_{e} E_{e}^{'}\; {\rm cos}^{2}(\theta_{e} /2)$
and $ y = 1 - (E_{e} / E_{e}^{'})\; {\rm sin^2}(\theta_{e} /2)$
 where $E_e$ is the incident beam energy.
 The 
scaling variable Bjorken-$x$ is related to these quantities via the square
of the centre of mass energy $s$: $x = Q^{2} / (ys)$.
Further reduction of photoproduction background  
and the removal of events in which a high energy
photon is radiated off 
the incoming electron 
are achieved by 
requiring      $\sum_j{(E_j-p_{z,j})} >35$ GeV \cite{f2pap},
with
the sum extending over all detected
particles $j$
in the event, and by the requirement $y> 0.1$, respectively.
The events are required to have an event vertex which is within 30 cm of the 
nominal vertex position.

All distributions of data shown in this paper have been corrected
bin-by-bin for 
detector effects, including geometrical acceptance,
for QED radiative effects, 
and for the detection efficiencies of
{\piz} mesons, 
    charged tracks and   jets~\cite{contreras} 
                respectively.                      
  The corrections were  
determined using events generated with the 
ARIADNE Monte Carlo program and 
 a full simulation
of the H1 detector response.

\subsection{\bf Forward \boldmath $\pi^0$ selection}

The {\piz}-mesons are measured using the dominant decay channel 
{\piz} $\rightarrow 2\gamma$. The {\piz} candidates are selected in
the region  $5^{\circ} < {\thpi} < 25^{\circ}$, where {\thpi} is the polar
angle of the produced {\piz}.   
Candidates are required to have an energy of 
{\xpi }={\epi}/$E_p >$ 0.01, with $E_p$ the proton beam energy, and 
a transverse momentum with respect to the beam axis, $p_{T_{\pi}}$, 
greater than  1 GeV. Here, $p_{T_{\pi}}$ is taken to be equal to
the transverse energy
 {\etpi} = {\epi} sin{\thpi}.
At the high {\piz} energies considered here, the two photons 
from the decay  cannot be
separated, but appear as one object (cluster) in the
calorimetric response. Photon induced showers are selected by
measuring the shower shapes of candidate clusters.  
The selection criteria are based on the compact nature 
of electromagnetic showers as opposed to those of hadronic origin.
The very fine segmentation of the LAr calorimeter in the forward region
(with a cell size
of $3.5 \times 3.5$ cm$^2$ and four-fold longitudinal segmentation for 
                   the electromagnetic calorimeter)
            makes possible a 
 detailed 
study of the transverse and longitudinal energy spread and the energy
of each cluster. The high particle density in the forward direction leads to
overlapping showers of electromagnetic and hadronic origin.
Candidates of this type are largely rejected by the following 
requirements.

A \piz\ candidate is  required to have more than $90 \%$ of
its  energy deposited in the electromagnetic part of the LAr
calorimeter. 
A ``hot'' core consisting of the most energetic
group of continuous electromagnetic
 calorimeter cells of a cluster, 
which must include the hottest
cell, is defined for each candidate~\cite{epsep}. More than 50\% of the
cluster energy is required to be deposited in this core.
The lateral spread of the shower
is quantified in terms of lateral shower moments calculated relative to 
the shower's principal axis~\cite{epsep} and required to be 
less than 4 cm~\cite{torsten}.
The longitudinal shower shape is used as a selection criterion via the
fraction of the shower's energy deposited in each layer of cells in the 
electromagnetic  part
of the calorimeter. The precise specifications
of   these layers can be
found in~\cite{epsep}. The value obtained by subtracting the fraction of
energy deposited in the fourth layer of cells from that in the second
layer as seen from the interaction point is required to be larger than
0.4, thereby selecting showers which start to develop close to the
calorimeter surface and are well contained in the 
electromagnetic part of the
calorimeter, as expected for showers of electromagnetic origin.
With this selection 1673    
{\piz} candidates are found  in the kinematic range  $5^{\circ} <
\theta_{\pi} < 25^{\circ}$, 
$x_{\pi} > 0.01$  and $p_{T_{\pi}} > 1$ GeV with a detection  efficiency
 above $39 \%$.

Monte Carlo studies using a detailed simulation of the H1 detector
predict that less than $20 \%$ of the \piz\ candidates
are due to other misidentified hadrons.   
                          These calculations also  show that less than $10 \%$ 
of the candidates originate from  secondary scattering of charged hadrons with passive material in
the forward region, where the amount of material between the interaction point and
the calorimeter surface is largest. This contamination is subtracted bin-by-bin.
Background from photoproduction has been studied with the PHOJET
 Monte Carlo model, following the procedure used in the
       measurement of $F_2$ \cite{f2pap}, and is found to make no
significant contribution to the distributions shown.

\subsection{Charged particle selection}

Charged particles are selected using the forward (FT) and central (CT)  
tracking chambers, in
the region  ~$5^{\circ} < {\thcp} < 25^{\circ}$, where {\thcp} is the polar
angle of the charged particle produced.   
Candidates are required to have an energy of {\xcp} = {\ecp}/$E_p >$ 0.01 and 
 a transverse momentum with respect to the beam axis,
$p_{T_{\mathrm{cp}}}$, greater than 1 GeV.
 Basic quality criteria are used to select well measured
tracks 
originating from the primary interaction point. Tracks measured in the 
CT are required to have at least 10 associated hits  
and a radial track length of over 10 cm. Selected tracks in the FT must 
have at least one planar segment confirmed   
by another segment reconstructed in either the planar or radial
chambers. Furthermore, the impact parameter 
in the plane transverse to the beamline 
of the track from the primary vertex is required to be less 
than 5 cm when measured at the vertex position in $z$.
  Where a track has been found in both the FT and CT, a 
combined momentum measurement 
is used. 
For this measurement, as for the {\piz}-mesons, the high particle
multiplicity in the forward
direction constitutes the main challenge. It gives rise to a large
flux of soft particles produced in secondary interactions in passive
material both in the FT and preceding it, 
                leading to a degradation of detection
efficiency and resolution.


After the application of these cuts 2251 charged particles 
were selected for this study. The efficiency of 
reconstructing a central track and associating it to the 
primary event vertex is over 95$\%$~\cite{pvm}. The equivalent 
forward track efficiency is  
over 40$\%$~\cite{FT} for tracks well contained within the FT. For 
tracks produced at low values of $\theta$  ($\theta < 10^\circ$), an
efficiency of 25$\%$ is obtained. 

The proportion  
of  selected charged particles that are produced in secondary
interactions in the passive material 
in and around the H1 detector is negligible for central
tracks. For the forward tracks it is 
 estimated by a detailed simulation of the 
H1 detector to be maximally
10\%. This contribution 
is subtracted bin-by-bin. The 
assumed material distributions within the detector in the forward 
region have been verified by measuring the  distribution of the 
rate of          $\gamma \rightarrow 
e^+e^-$  conversion pairs throughout the forward region of H1.

\subsection{\bf Forward jet selection}

 The  region in which the forward jet measurements in this paper are made is  
chosen such that  the phase space for jet production arising via 
 DGLAP evolution is suppressed compared to that available 
  in the case of BFKL evolution. This is achieved  by requiring
 $\kjet^2\approx Q^2$, where  $\kjet^2$ is the transverse
momentum squared of the forward partonic jet. Due to the strong ordering
in $k_T^2$, this  
leaves little room for DGLAP evolution.  In addition,
the momentum fraction of the jet, $\xjet = \Ejet/E_p$,
should be as large as 
possible, whereas the momentum fraction \xB of the quark struck by the
virtual photon should be small in order to maximize the phase space
for jet production from BFKL evolution 
which is governed by the ratio $\xjet/x$.
Here $\Ejet$ and $E_p$ are the energies of the jet and the incoming
proton respectively.
The rate of events at small \xB with a jet at high $\xjet$ should 
 be much  higher for the BFKL than for the DGLAP scenario, as 
was shown  in~\cite{adrbartels}.
\begin{figure}[tb] \centering
\epsfig{file=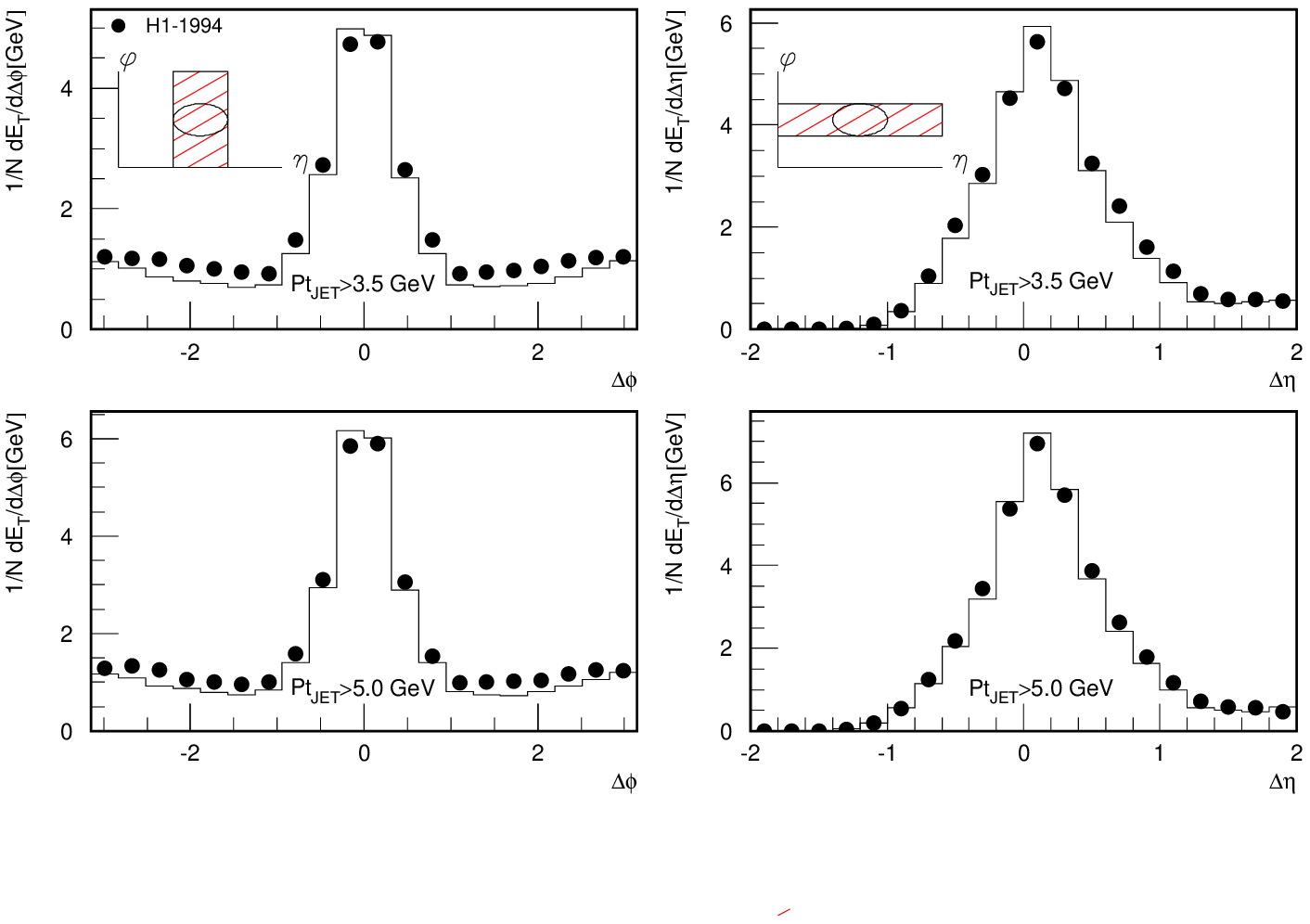,width=150mm}
\caption{
The average transverse  energy flow around  the forward jet
axis over all events is shown for two different minimum $\pjet$
values
as a function of $\Delta \phi$ (left ),  integrated over
$|\Delta \eta|< 1$ and as a function of $\Delta\eta$, (right),
    integrated over $|\Delta \phi| < 1$ rad, as shown by the insets in
the top figures.
Here $\Delta \eta $
and $\Delta \phi$ are measured with respect to the
reconstructed  jet axis. Also shown is the DJANGO expectation
 with full H1 detector simulation.}
\label{profiles}
\label{SLOPE}
\end{figure}

A simple cone algorithm~\cite{contreras}
 is used to find jets in the H1 LAr calorimeter, 
requiring an $E_T$ larger than 
3.5~GeV in a cone of radius 
$ R = \sqrt{\Delta\eta^2 + \Delta\phi^2} = 1.0$ in the space of
 pseudo-rapidity $\eta$ and azimuthal angle $\phi$
in the HERA frame of reference. This jet algorithm was found to 
give the best correspondence between jets at the detector
and hadron level for this type of analysis~\cite{kuerz}.
 We have measured the cross-section for 
events which have a ``forward'' jet defined by 
$\xjet=\Ejet/E_p\!>\!0.035$, $0.5\!<\! \pjet^2/Q^2\!<\!2$, 
$7^{\circ} < \thjet< 20^{\circ}$ and $\pjet> 3.5$ and 5  GeV, 
where $\pjet$ is the transverse momentum of the jet, taken to be 
equal to $E_T$.
A total of 1945 events contained at least one jet satisfying these criteria. 

The transverse energy flow around the forward jet axis,
averaged over all selected events,
is shown versus $\Delta\eta$ and $\Delta\phi$  with 
respect to the reconstructed jet axis in Fig.~\ref{profiles}, 
for various 
$\pjet$ ranges. 
Distinct jet profiles are observed which are reasonably
     described by the DJANGO model with full detector simulation.  

Backgrounds from photoproduction and radiative events (for which the 
current jet is boosted into the forward region) have been studied with
Monte Carlo and with real data. 
The proportion of radiative events was reduced by 
requiring a  cluster with energy larger than 0.5 GeV and pseudorapidity
 between $-1.3$  and 1.5 units of rapidity  away from the forward jet axis,  
 to tag the hadronic activity of the current quark.
A fraction  of the background  events can be 
experimentally measured (tagged) using the 
detectors of the luminosity system: for radiative events  
with a photon emitted collinearly with the incident electron, the photon 
can be detected  with the photon tagger;  for photoproduction events
the electron which is scattered through a
 small angle can be 
detected in the electron tagger. Using  data samples  with tagged
events it is possible
 to check the Monte Carlo based estimates of the photoproduction 
background.  This is found to be  less than  3\%  and
subtracted from the 
data presented below.

The forward jet analysis has also been repeated
 using an unfolding procedure~\cite{ewelina}
based on the Bayes unfolding method~\cite{bayes}.
The results agree very well with the bin-to-bin corrected  data.

\begin{figure}[t]
\epsfig{file=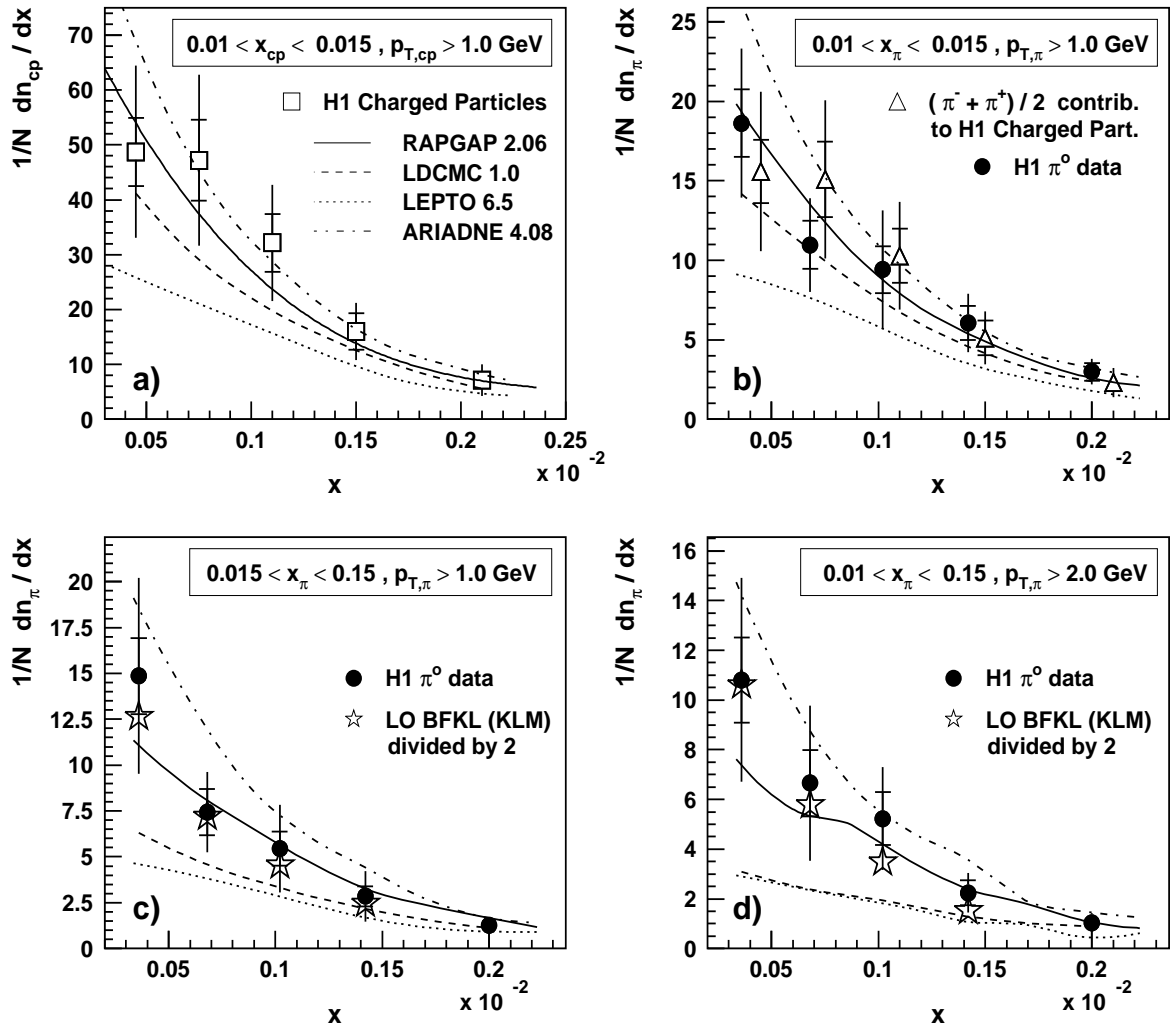,height=15cm,width=9cm,bbllx=30pt,bblly=185pt,bburx=320pt,bbury=650pt}
\caption{The single particle 
spectra in Bjorken-$x$ are shown for charged particles
  and {\piz}-mesons produced in the polar angle range 
$5^{\circ} < \theta < 25^{\circ}$. In the upper right plot
  ARIADNE with JETSET was used to calculate the contribution of charged
pions to the
  charged particle measurement which makes possible,
 when divided by two, a direct
  consistency check of the two measurements. For comparison, four
  different Monte Carlo models are overlaid, as well 
as an analytical calculation labelled BFKL(KLM)
  based on \protect \cite{kwiecinski}. 
  $n_{\pi}$ is the
  number of {\piz}-mesons and  $N$ is the number of DIS events that
  fall into the specified kinematic range. The full errors are the
  quadratic sum of the statistical (inner error bars) and systematic
  uncertainties.}   
\label{pi0spec}
\end{figure}

\section{Results}

 In Fig.~\ref{pi0spec},  single particle spectra are presented as a function
of Bjorken-$x$  for 
{\piz}-mesons and charged particles. The {\piz}-mesons are shown for  
$5^{\circ} < \theta_{\pi} < 25^{\circ}$ in two bins of 
$x_{\pi}$ (0.01 $<$ $x_{\pi}$ $<$ 0.015 and 
0.015 $<$ $x_{\pi}$ $<$ 0.15) for $p_{T_{\pi}} > 1$ GeV and for the higher
threshold of $p_{T_{\pi}} > 2$ GeV in the $x_{\pi}$ range 
0.01 $<$ $x_{\pi}$ $<$ 0.15.
All distributions are normalized to the number of DIS events, $N$, 
 in the kinematic region specified in section~\ref{sec:datasel} and with
the additional requirement that 
 Bjorken-$x$ is in the range  $0.0002 < x < 0.00235$.  
The full errors are the quadratic sum of the
statistical (inner error bars) and systematic uncertainties. 
For the {\piz} measurements the systematic errors contain the model 
dependence of the correction, estimated from  the difference in correction 
factors obtained using LEPTO
and 
ARIADNE, which leads to  errors of between 20-30\%;  a
4\%
 variation of the electromagnetic energy scale in the forward region gives 
a 5\% uncertainty;
the variation of \piz\ 
selection cuts contributes  5-10\% systematic error; and
the variation of \piz\ 
acceptance also produces a  5-10\% uncertainty. 

The dominant source of 
systematic uncertainty for the charged particle measurements stems from the 
dependence of the FT efficiency on the
forward particle multiplicity, which leads to a large systematic 
error due to the model dependence of the correction procedure. This gives  
typical point-to-point errors of between 10\% and 25\% in the 
analysis presented here. These numbers were  
evaluated using simulated LEPTO and ARIADNE 
events. Discrepancies between the simulation and the 
FT detector response give rise to further point-to-point systematic 
errors of about 15$\%$. Other sources of systematic effects, 
such as the BEMC energy scale uncertainty and photoproduction background are
included in the  errors but are small compared to those  
quoted above.

The data in Fig.~\ref{pi0spec}
clearly show that the production of {\piz}-mesons with high 
$p_{T}$ increases     towards small Bjorken-$x$.
The binning was chosen to match the $x$ resolution. Only
for the lower  bin in $x_{\pi}$ and $p_{T}$  is a measurement of charged
particles                                        
available. To check the consistency of both measurements the
 charged particles results have been converted to 
                charged pions spectra  using the charged pion fraction       
predicted by     ARIADNE with JETSET,             which describes
the charged particle data well. 
From isospin symmetry the neutral pion spectrum is expected to 
be half of the total charged pion spectrum.
The resulting charged pion spectrum, divided by a factor two, is 
overlaid on Fig.~\ref{pi0spec}b, 
and is found to be                             in good agreement with the
{\piz} measurement.
The model predictions shown
will be discussed below.

\begin{figure}[htb] \centering
\epsfig{file=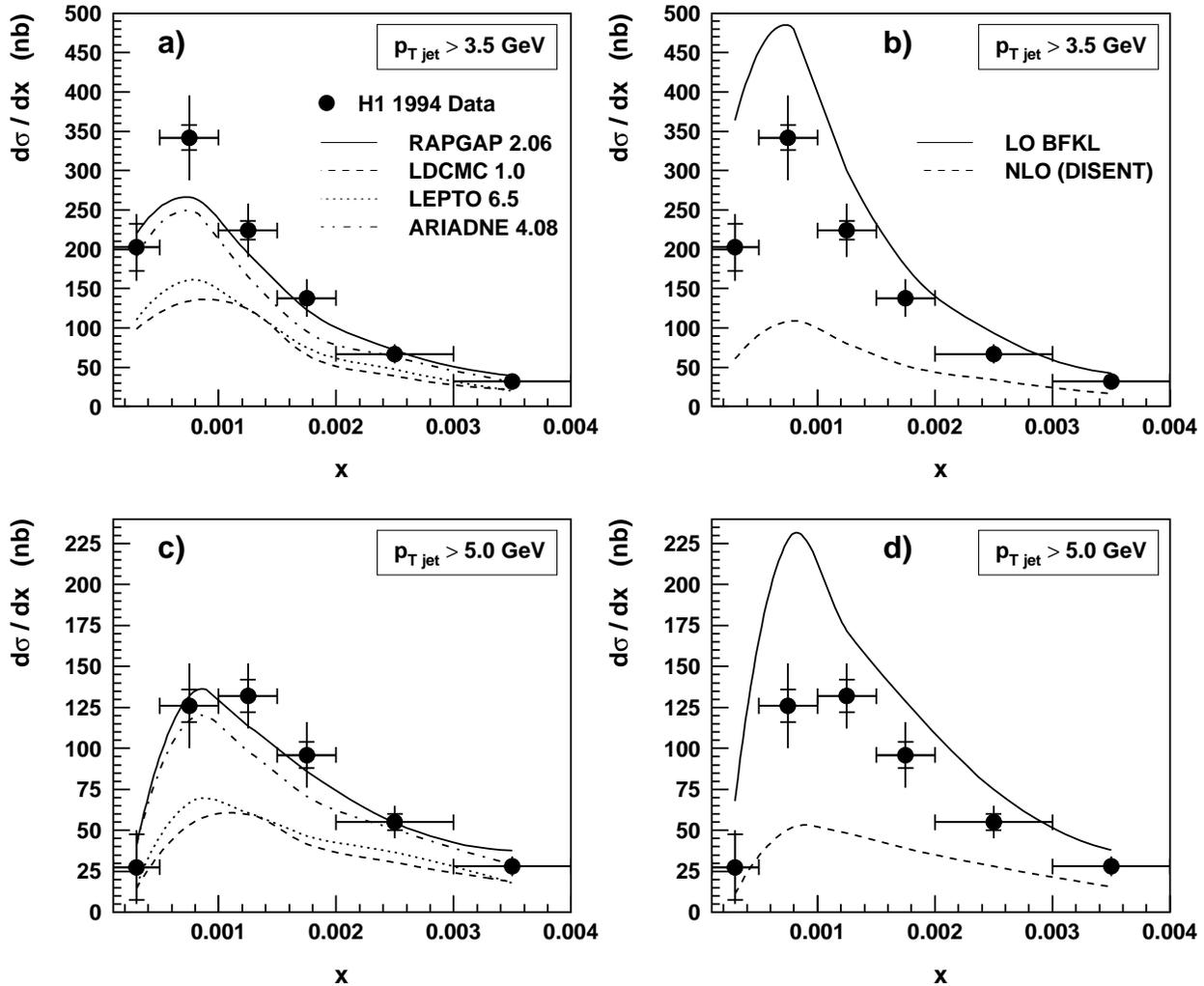,width=150mm,bbllx=50pt,bblly=180pt,bburx=520pt,bbury=650pt}
\caption{ The
forward jet cross-section  as a function of Bjorken-$x$ for two
$\pjet$ cuts: 3.5 GeV and 5.0 GeV.
 The errors shown are the statistical and systematic uncertainties
added in quadrature. In (a) and (c) the curves are model calculations
(full line RAPGAP, dashed line LDCMC, dotted line LEPTO and the
dashed dotted line ARIADNE). In (b) (d) the full lines are
analytic LO BFKL calculations at the
parton level (no jet algorithm was applied), the dashed lines $O(\alpha_s^2)$
calculations using DISENT with a cone algorithm applied.
}
\label{fwdjet}
\end{figure}

The forward jet data for $\pjet > 3.5 $ GeV  
and  $\pjet > 5$ GeV are shown as a 
function
of Bjorken-$x$ in the range $0.0001 < x< 0.004$ in Fig.~\ref{fwdjet}. 
In the case
 that there are two jets in the event which satisfy the jet selection 
criteria (approximately 1\% of the forward jet events, 
  see below), the jet with the largest
$\pjet$ was taken. 
It was shown that with the HERA kinematics and 
detector  constraints, which determine the smallest reachable
angle for $\thjet$, the largest differences between 
BFKL and fixed order calculations  are expected
for low values of $\pjet$, namely around 3.5 GeV \cite{wsderoeck}. 
Reducing this value further
 leads to excessively large hadronisation corrections as predicted 
by ARIADNE  and
dubious jet
selections. However, to study the sensitivity of this study 
        to  
 increasing $\pjet$,  
cross-sections with  $\pjet > 5$ GeV have also been measured.
The data points in Fig.~\ref{fwdjet}
present the cross-section as measured 
in the given $x$ bin for the kinematic and jet cuts defined
above.
The bin purity, i.e. the fraction  of events in a bin in $x$
with a jet at the detector level which also have a jet at the hadron level 
in that bin,  
is between 25\% and 50\% for the lower $\pjet$ threshold 
value and somewhat larger
for the higher $\pjet$ threshold value.

The data shown in Fig.~\ref{fwdjet}
clearly rise with decreasing $x$, except for the smallest
$x$ bin, for which the rise is hidden due to the narrow kinematic region
left with the present  cuts.
The total cross-section for forward jets 
with $\pjet> 3.5$ GeV in the defined kinematic range
is $531\pm17(stat.)^{+82}_{-89}(syst.)$ pb.

\begin{figure}[htb] \centering  
\epsfig{file=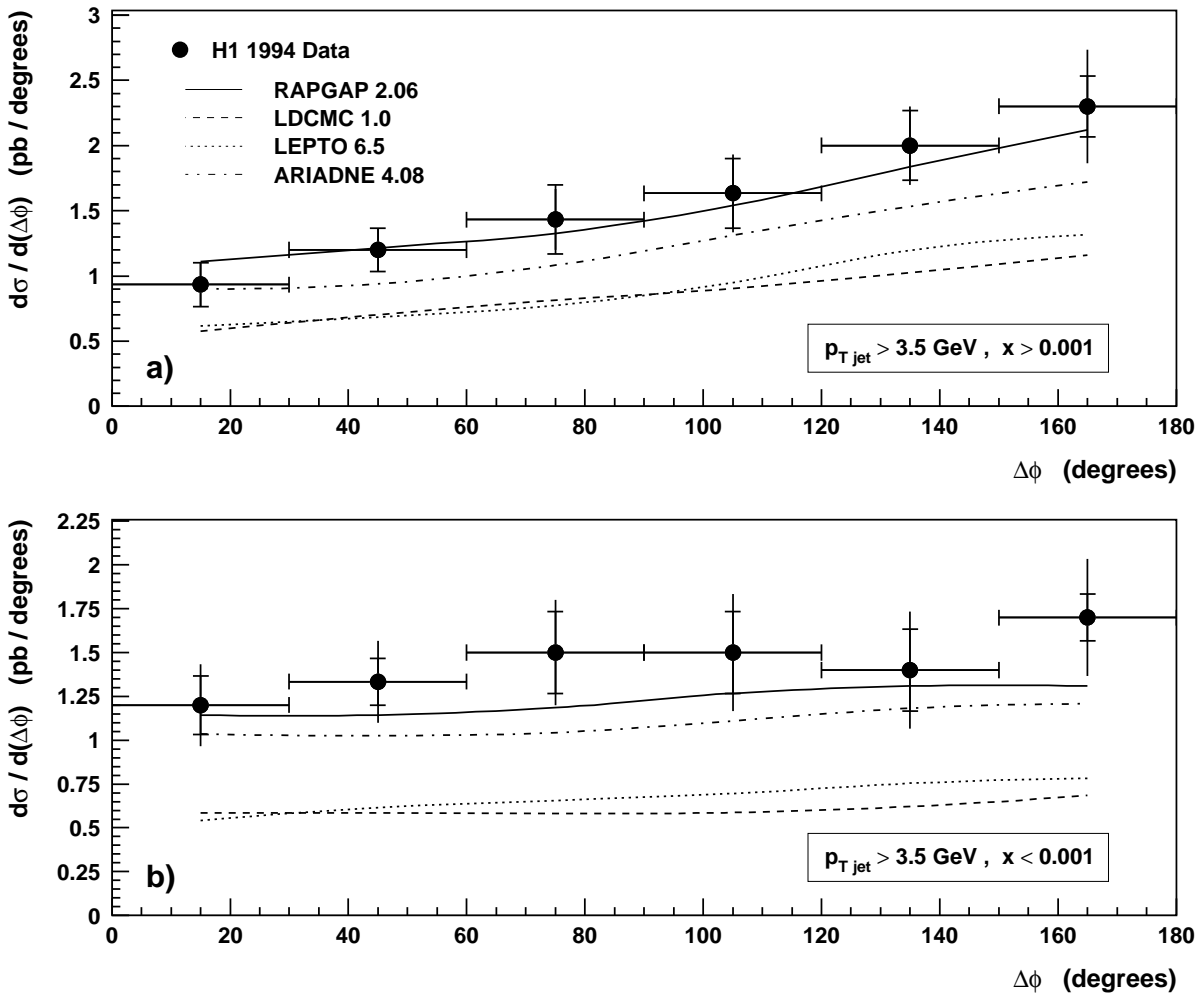,width=180mm,bbllx=20pt,bblly=170pt,bburx=600pt,bbury=670pt} 
\caption{ The azimuthal correlation $\Delta \phi$
(in degrees) between the jet and the
scattered electron, for two different $x$ regions.
The curves are model calculations
(full line RAPGAP, dashed line LDCMC, dotted line LEPTO and the
dashed dotted line ARIADNE).}
\label{angle}
\end{figure}

The systematic errors on the forward jet data points shown in
Fig.~\ref{fwdjet} and in Table~\ref{tab001} are
dominated by the 
 4\% uncertainty of the hadronic energy scale, leading to shifts in 
the jet rates of 12\%. Another important systematic contribution
results from the dependence on the model used to correct the data and is
up to 7\% (the data shown are corrected with the ARIADNE model). The 
energy scale of the BEMC
 affects the cross-section by 3-4\% in Figs.~\ref{fwdjet} and
~\ref{angle}. Other systematic effects have been found to be small compared 
to the above.

\begin{table}
\begin{center}
\begin{tabular}[h]{|c|c|}
\hline
Event Sample  & Forward jet cross-section [pb]  \\
\hline
 H1 1994 Data        &  $531\pm17(stat.)^{+82}_{-89}(syst.)$ \\
\hline
 ARIADNE 4.08        &  425 \\
\hline
 RAPGAP 2.06        &  491 \\
\hline
 LEPTO 6.5       &  259 \\
\hline
 LDCMC 1.0       &  262 \\
\hline
\end{tabular}
\caption{  Cross-sections for H1 data and QCD calculations for
inclusive forward
 jets satisfying
$\xjet >\!0.035$, $0.5\!<\! \pjet^2/Q^2\!<\!2$,
$7^{\circ} < \thjet< 20^{\circ}$ and $\pjet> 3.5$ GeV
in the kinematic range  $E_{e}^{'} > 11 $ GeV,
$160^{\circ} < \theta_{e} < 173^{\circ}$ and $y > 0.1$. }

  \label{tab001}
\end{center}
\end{table}

        The Monte Carlo models described above are
compared to the data in Fig.~\ref{fwdjet}. 
The data rise faster with decreasing $x$
than LDCMC and LEPTO, which are 
based on the CCFM equation and LO DGLAP dynamics respectively. The
prediction 
of ARIADNE is slightly below the jet data and
is slightly above the single particle data. ARIADNE does however 
reproduce the rise towards low $x$. Unlike LEPTO,
this model does not include
a $k_T$ ordered cascade and it has been suggested that these calculations  
  will be  close to the BFKL prediction~\cite{bfklcdm}. It has however  
 also been hypothesised~\cite{rathman} that the sizeable QCD radiation
predicted by this model in the forward region stems not from the lack of
a $k_T$ ordered  cascade but from an unconventional enhancement of
QCD-Compton processes at low $x$ in which a hard gluon is radiated
after the parton-photon interaction. 
The ARIADNE predictions also depend on the photon and 
target size
extension parameters used in the calculations\footnote{The default values 
PARA(10)=1.0 and 
PARA(14)=1.0 have been used.
Changing these parameters to 1.5 (0.5) reduces (increases) the 
cross-sections by approximately 15-30\%}.

The RAPGAP model, which is based on LO DGLAP dynamics but has 
in addition a contribution from resolved photons, gives a good
description of both the single particle and jet data. The RAPGAP
Monte Carlo allows the study of the scale dependence of the result.
Sensible variations of the choice of scale used\footnote{
The scale was changed from $p^2_T+Q^2$ to $4p^2_T+Q^2$.} in the 
 leading order calculation were found to reduce the resolved  
 photon cross-section calculations by up to 40\%. The direct 
cross-section is more stable with respect 
to the choice of scale and changes by 
less than 10\%. 
 A comparison of the integrated forward jet cross-section 
 for data and for these Monte Carlo models is given in Table~\ref{tab001}.
The Monte Carlo cross-sections are based on event samples
with numbers of events which are a factor
2 to 3  larger than those in the data samples.

 The forward jet data, shown in Fig.~\ref{fwdjet}
are  also compared with recent
 analytical LO BFKL calculations~\cite{adrbartels} and a fixed order
$O(\alpha_s^2)$ jet calculation
 at the parton level. No jet algorithm is used for the BFKL 
calculation while a cone algorithm~\cite{cdfcone} has been used for the
fixed order calculation. 
The BFKL calculation is above  the
 measurements  for the data with both $\pjet$ thresholds of 3.5 and 5 GeV.
The fixed order prediction for direct interactions of the photon
with the proton is well below the 
data and very close to the so called 
 ``Born'' calculation in ~\cite{adrbartels}, for which  no  gluon
emissions are allowed in the BFKL ladder. 
The latter    corresponds to the quark box 
diagram seen by the photon, and one single gluon at the low end of the 
diagram as is explained in~\cite{adrbartels}. 
Variations of the fixed order cross section predictions of about 40\%
are found when the scale ($Q^2$) is varied by a factor of four. 

Hadronisation corrections and  kinematic constraints 
(for the BFKL 
calculations) may change the  predictions made at the parton level.
The former are unfortunately strongly model dependent and at the lowest
$x$ values  
LEPTO and ARIADNE predict a change  of the jet 
cross-section due to hadronisation 
by   factors (number of jets at parton level/
number of jets at hadron level) of 1.0 and 0.7 for $\pjet > 3.5$ GeV,  
and 1.5 and 1.1 for   
$\pjet> 5$ GeV respectively. The kinematic constraints are expected 
to lower the predicted cross-sections~\cite{stirling}.  
It should also be noted that the BFKL calculations contain
a typical power growth of the cross-section with $x$ with
a power of about 0.5. Recently, next-to-leading order  
BFKL calculations 
 for the inclusive cross-sections have become available~\cite{fadin} 
which demonstrate large corrections to this power, substantially
reducing it. 
The question of how these large corrections will change the
  numerical BFKL predictions for the HERA forward jets 
  cross-sections has not yet been addressed; in fact a
  recent paper~\cite{Ross} argues that the next-to-leading
order reduction of the BFKL
  exponent might be somewhat smaller than naively expected. Furthermore,
the next-to-next-to-leading order  corrections may also be large.  
In view of these uncertainties, the interpretation of possible
  deviations of experimental data from fixed order matrix element
  calculations as being due to BFKL effects has become more difficult.

BFKL calculations folded with fragmentation 
 functions are also available for the highest 
 momentum {\piz} bins. 
This prediction is 
a factor of two above the data although the shape is in 
well described. This is shown in Fig.~\ref{pi0spec} in
which 
the prediction, divided by two~\cite{martin}, is compared to the data.
 Despite this normalisation discrepancy the calculations  
 are seen to describe well the $x$ dependence of the {\piz} measurements.

It has been suggested that a 
study of the azimuthal correlation 
$\Delta \phi$ between the  scattered electron and the forward
jet~\cite{adrbartels} may be a sensitive discriminant between the BFKL and
DGLAP evolution schemes. A stronger decorrelation at low $x$ is expected in the BFKL case 
than in that  of 
DGLAP due to 
the dilution of the original correlation resulting from multi-gluon 
emission. The data are shown in 
Fig.~\ref{angle} for the smaller and higher $x$ regions.
The data exhibit the expected levelling of the distribution when 
$x$ decreases although all models follow this
tendency, suggesting that this is not a very sensitive variable in this
kinematic range.

Of the 1945 forward jet 
events selected, 52 have a second jet which fulfill the
forward jet
criteria. This corresponds to a total  cross-section of
$6.0\pm0.8(stat.){\pm 3.2}(syst.)$
pb,
or roughly 1\% of the forward jet cross-section. 
\begin{table}
\begin{center}
\begin{tabular}[h]{|c|c|}
\hline
Event Sample  & Forward di-jet cross-section (pb) \\
\hline
 H1 1994 Data        &  $6.0\pm{0.8}(stat.)^{+3.2}_{-3.1}(syst.)$ \\
\hline
 ARIADNE 4.08       &  6.7 \\
\hline
 RAPGAP 2.06       &  6.0 \\
\hline
 LEPTO 6.5       &  1.8 \\
\hline
 LDCMC 1.0       &  2.1 \\
\hline
\end{tabular}
\caption{  Cross-sections for H1 data and different model calculations for
 forward
di-jet events, each satisfying
$\xjet>\!0.035$, $0.5\!<\! \pjet^2/Q^2\!<\!2$,
$7^{\circ} < \thjet< 20^{\circ}$ and $\pjet> 3.5$ GeV
in the kinematic range  $E_{e}^{'} > 11 $ GeV ,
$160^{\circ} < \theta_{e} < 173^{\circ}$ and $y > 0.1$. }
  \label{tab002}
\end{center}
\end{table}
 Recent LO BFKL  numerical studies 
at the parton level~\cite{alan} predict that 
 3\% of events will have two or more forward jets. We have also checked 
that this discrepancy between the data and the BFKL calculation
 remains if the ratio of $Q^2$ to $\pjet^2$ is relaxed to
$0.5\!<\! \pjet^2/Q^2\!<\!5$.
However, the BFKL calculations, as before,  include neither the effects of    
hadronisation  nor those of jet algorithms. 
 Table~\ref{tab002} compares the 
forward di-jet cross-section in data with the QCD models.
ARIADNE and RAPGAP give di-jet rates in agreement with the data, while the
other models predict lower values and, owing to the current precision of
the measurement,  cannot yet be excluded.

\section{Summary}
\noindent

   Measurements of the cross sections for the production of  forward jets
and
high transverse momentum single 
particles in low $x$ DIS have been presented. Models whose dynamics are 
based on different schemes for QCD evolution have been compared with 
the results. The conclusions using the forward jet and the single particle 
measurements are in agreement. 
 
Models implementing the traditional DGLAP evolution 
and including only direct photon interactions grossly underestimate
the amount of perturbative radiation required. Also a  $O(\alpha_s^2)$
 jet calculation predicts less jet production.
Models which include also the resolved photon component
successfully describe all the data presented.
BFKL
calculations at leading order describe the strong rise of the
forward single jet and particle cross-sections at low $x$. 
 While BFKL 
effects are expected to be prominent in this low $x$ domain complete 
BFKL calculations have still to be performed at 
next-to-leading order before the data can be interpreted in this light. 

Predictions of a model based on an implementation of 
CCFM evolution, which should smoothly interpolate 
between the DGLAP and BFKL regimes, give a poor description of all
measured distributions. Calculations implementing the  
Colour Dipole Model provide sufficient QCD radiation to match the 
data.

%

\section*{Acknowledgements}

We are grateful to the HERA machine group whose outstanding
efforts have made and continue to make this experiment possible. 
We thank
the engineers and technicians for their work in constructing and now
maintaining the H1 detector, our funding agencies for 
financial support, the
DESY technical staff for continual assistance, 
and the DESY directorate for the
hospitality which they extend to the non-DESY 
members of the collaboration.
We would like to thank J. Bartels, J. Kwiecinski,  S. Lang and W. J. Stirling
for useful discussions.


\newpage


\begin{thebibliography}{99}



\bibitem{mueller}
  A. H. Mueller, Nucl. Phys.  B (Proc. Suppl.)  18C (1990) 125; 
  J. Phys. G17 (1991) 1443. 

\bibitem{dhotref}
  J. Kwieci\'{n}ski, A.D. Martin, P.J. Sutton, Phys. Rev. D46 (1992) 921.


\bibitem{hotref}
  J. Bartels, A. De Roeck, M. Loewe, Z. Phys. { C54} (1992) 635;\\
  W. K. Tang, Phys. Lett. B278 (1992) 363.

\bibitem{forwjets} H1 Collab., S. Aid et al. Phys. Lett. B356 (1995) 118.

\bibitem{fjzeus} ZEUS Collab., J. Breitweg et al.,
DESY-98-050, hep-ex/9805016 (1998).

\bibitem{kuhlen} M. Kuhlen, Phys. Lett. B382 (1996) 441.

\bibitem{charged} H1 Collab., C. Adloff et al., Nucl. Phys. B485 (1997) 3.

\bibitem{dglap}
  Yu. L. Dokshitzer, Sov. Phys. JETP 46 (1977) 641; \\
  V. N. Gribov and L. N. Lipatov, Sov. J. Nucl. Phys. 15 (1972) 438 and 675; \\
  G. Altarelli and G. Parisi, Nucl. Phys. 126 (1977) 297.

\bibitem{bfkl}
  E. A. Kuraev, 
L. N. Lipatov and V. S. Fadin, Sov. Phys. JETP 45 (1972) 199; \\
  Y. Y. Balitsky and L. N. Lipatov, Sov. J. Nucl. Phys. 28 (1978) 822. 

\bibitem{ccfm} M. Ciafaloni, Nucl. Phys. B296 (1988)  49;\\
S. Catani, F. Fiorani and G. Marchesini, Phys. Let. 234B (1990) 339;\\
S. Catani, F. Fiorani and G. Marchesini, Nucl. Phys. B336 (1990) 18.


\bibitem{muellercarg}
  A. H. Mueller, Columbia preprint CU-TP-658 (1994) and 
 Cargese Summer Inst. (1994) 87.



\bibitem{bartels}
  J. Bartels and H. Lotter,  Phys. Lett. B309 (1993) 400.






\bibitem{jung} H. Jung, L. J\"onsson, H. K\"uster, hep-ph/9805396 (1998),
DESY-98-51.

\bibitem{lepto} 
  G. Ingelman,       
  Proc. of the HERA workshop, eds. W.~Buchm\"uller and
  G.~Ingelman, Hamburg (1992) vol. 3, p. 1366.
\bibitem{tania}
 H1 Collab., C. Adloff et al. Phys. Lett. B415 (1997) 418.

\bibitem{h1dijets} 
H1 Collab., C. Adloff et al., DESY 98-076, submitted to Europ. Phys.
J. C.

\bibitem{rapgap} H. Jung, Comp. Phys. Comm. 86 (1995) 147; 
(for update see http://www-h1.desy.de/\string~jung/rapgap/rapgap.html).

\bibitem{SAS1D}  G. A. Schuler, T. Sjostrand, Phys. Lett. B376 (1996) 193.

\bibitem{ariadne}
{L. L\"onnblad, Comp. Phys. Comm.  71 (1992) 15.}


\bibitem{dipole}
 G. Gustafson, Phys. Lett. B175 (1986) 453; \\
 B. Andersson, G. Gustafson, L. L\"onnblad and  U.  Pettersson,
 Z. Phys. C43 (1989) 625;\\
G. Gustafson, U. Pettersson, Nucl. Phys. B306 (1988) 746. 

\bibitem{bfklcdm}
 L. L\"onnblad, Z. Phys. C65 (1995) 285; \\
 A. H. Mueller, Nucl. Phys. B415 (1994) 373;\\
 L. L\"onnblad, Z. Phys. C70 (1996) 107.     


\bibitem{ldc}
B. Andersson, G. Gustafson, J. Samuelsson,  Nucl. Phys. B467 (1996) 
443.\\
B. Andersson, G. Gustafson, H. Kharraziha, J. Samuelsson,
 Z. Phys. C71  (1996) 613.

\bibitem{Lonnblad-ldc}
H. Kharraziha, L. L\"onnblad, hep-ph/9709424 (1997).

\bibitem{string}
 T. Sj\"ostrand, Comp. Phys. Comm. 39 (1986) 347; \\
 T. Sj\"ostrand and M. Bengtsson, Comp. Phys. Comm. 43 (1987) 367; \\
 T. Sj\"ostrand, CERN-TH-6488-92 (1992). 

\bibitem{GRV}  M. Gl\"uck, E. Reya and A. Vogt, Z. Phys {C67} (1995)
433.

\bibitem{django} G. A. Schuler and H. Spiesberger, Proc. of the HERA
Workshop, Vol 3. ,
Eds. W.~Buchm\"uller and G. Ingelman, DESY 1992, 1419.



\bibitem{adrbartels} J. Bartels et al.,  Phys. Lett. B384 (1996) 300.

\bibitem{kwiecinski} J. Kwiecinski, S. C. Lang and A. D. Martin, 
hep-ph/9707240;\\
J. Kwiecinski, S. C. Lang and  A. D. Martin,
Phys. Rev. D55 (1997) 1273.


\bibitem{disent} S. Catani, M.H. Seymour,
Nucl. Phys. B485 (1997) 291, Erratum-ibid. B510 (1997) 503.


\bibitem{phojet}
{ R. Engel, Proceedings of the XXIXth Rencontre de Moriond (1994) 321.}

\bibitem{h1nim}
H1 Collab., I. Abt et al.,
Nucl. Instr. and Methods A386 (1997) 310. \\
H1 Collab., I. Abt et al.,
Nucl. Instr. and Methods A386 (1997) 348.


\bibitem{larc}
  H1 Calorimeter Group, B. Andrieu et al.,
  Nucl. Instr. and Meth. A336 (1993) 460.

\bibitem{h1pi}
  H1 Calorimeter Group, B. Andrieu et al.,
  Nucl. Instr. and Meth. A336 (1993) 499.

\bibitem{ewelina} E. M. Lobodzinska, PhD. Thesis, Krakow University,
1997 (unpublished).

\bibitem{f2pap} 
  H1 Collab., S. Aid et al., Nucl. Phys. B470 (1996) 3.













  

























\bibitem{bemcnim} 
  H1 BEMC group, J. Ban et al., Nucl. Inst. and Meth. A372 (1996) 399.

\bibitem{FT}
{S. Burke et al., Nucl. Instr. and Meth.  A373 (1996) 227. } 


\bibitem{contreras} J. G. Contreras Nuno, PhD. Thesis, Dortmund
University
1997 (unpublished).



\bibitem{epsep}
{ H1 Calorimeter Group, B. Andrieu et al., Nucl. Instr. and Meth. A344 (1994)}
492.

\bibitem{torsten} T. Wengler, PhD. Thesis, Heidelberg University,
to appear.


\bibitem{pvm}
{H1 Collab., S. Aid et al., Z. Phys. C72 (1996) 573}.


\bibitem{kuerz} J. Kurzh\"ofer,  PhD. Thesis, Dortmund
University
1995 (unpublished).
\bibitem{bayes} G. D. Agostini, Nucl. Instr. Meth. A362 (1995) 487.

\bibitem{wsderoeck} J. Bartels, A. De Roeck and M. W\"usthoff, 
Proceedings of the workshop Future Physics at HERA, Eds.
G. Ingelman, A. De Roeck, R. Klanner, p.~598 (1996).

\bibitem{rathman} J. Rathsman, Phys. Lett. B393 (1997) 181.


\bibitem{cdfcone} CDF Collab., F. Abe et al., Phys. Rev. D45 (1992) 1448.




\bibitem{stirling} W. J. Stirling, talk at the workshop on Deep Inelastic
Scattering, DIS98, Brussels, to appear in proceedings.


\bibitem{fadin} V.S.  Fadin and L. N. Lipatov, 
Phys. Lett. B429 (1998) 127.
\bibitem{Ross} D. A. Ross, hep-ph/9804332 (1998), SHEP-98-06.

\bibitem{martin} 
  A. Martin and S. Lang, private communication.

\bibitem{alan} J. Kwiecinski, C. A. M. Lewis and  A. D. Martin,
Phys. Rev. D57 (1998) 496.



















\end{thebibliography}
\end{document}